\documentclass[aps,prl,twocolumn,groupedaddress,showpacs]{revtex4}
\usepackage{graphicx}
\usepackage{dcolumn}
\begin{document}

\title{Orientational Defects in Ice Ih: An Interpretation of Electrical Conductivity 
Measurements}
\author{Maurice de Koning$^1$, Alex Antonelli$^1$, Antonio J. R. da Silva$^2$ and A. 
Fazzio$^2$}

\affiliation{$^1$Instituto de F\'{\i}sica Gleb Wataghin,Universidade Estadual de 
Campinas, Unicamp, 13083-970, Campinas, S\~ao Paulo, Brazil}
\affiliation{$^2$Instituto de F\'isica, Universidade de S\~ao Paulo, Caixa Postal 
66318, 05315-970, S\~ao Paulo, S\~ao Paulo,Brazil}

\begin{abstract}
We present a first-principles study of the structure and energetics of Bjerrum
defects in ice Ih and compare the results to experimental electrical conductivity
data. While the DFT result for the activation energy is in good agreement with
experiment, we find that its two components have quite different values. Aside from 
providing new insight into the fundamental
parameters of the microscopic electrical theory of ice, our results suggest the
activity of traps in doped ice in the temperature regime typically assumed to be
controlled by the free migration of L defects.
\end{abstract}


\date{\today}

\maketitle

While the isolated water molecule is one of the simplest in Nature, the condensed 
phases of ${\rm H_2O}$ reveal many complex features that still elude complete 
understanding~\cite{Kuo2004,Murray2005}. An example is one of the most abundant 
crystalline solids on Earth, the proton-disordered hexagonal ice Ih, for which 
several aspects of structure-properties relationship have yet to be 
clarified~\cite{Petrenko}. 

An important issue concerns the role of crystal defects in the peculiar electrical 
properties of ice Ih. When an electric field is applied to an ice specimen, it 
becomes polarized by the thermally activated reorientation of the molecular dipoles. 
To explain the molecular origin of this phenomenon, Bjerrum~\cite{Bjerrum1952} 
postulated the existence of orientational defects that represent local disruptions of 
the hydrogen-bond network of ice Ih. While the conceptual picture of these Bjerrum 
defects is now well established,~\cite{Petrenko} a quantitative understanding of 
their structure and energetics is still lacking, rendering a direct interpretation of 
experimental electrical conductivity data difficult. 

Recent atomistic studies~\cite{Hassan1992,Podeszwa1999,Grishina2004} based on 
empirical potentials have provided important qualitative insight into the structure 
and dynamics of Bjerrum defects, but have not yet attempted to make direct contact 
with experimental conductivity data. Furthermore, the few {\em ab initio} 
studies~\cite{Newton1983,Plummer1992} in  the literature involved clusters that are 
too small to reliably capture the properties of a defect embedded in bulk crystal. In 
this Letter, we present an {\em ab initio} study of the structure and energetics of 
Bjerrum defects in ice Ih using a large supercell. Moreover, based on Jaccard's 
defect-based microscopic electrical theory of ice~\cite{Petrenko,Jaccard1959}, we 
interpret the results in terms of experimental electrical conductivity data for doped 
ice Ih. 

\begin{figure}
\includegraphics*[scale=0.7]{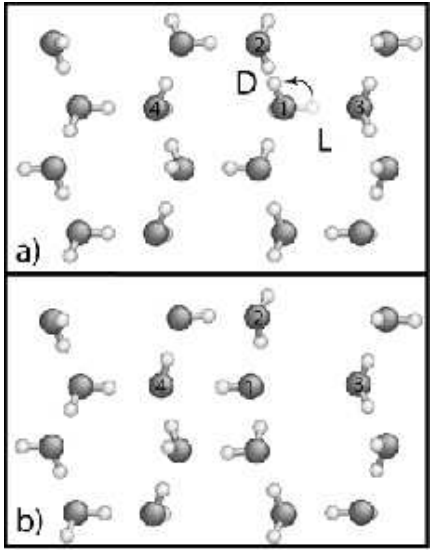}
\caption{\label{fig1} a) Schematic picture of the formation of a Bjerrum defect pair 
from the defect-free ice Ih structure. The defect pair is created by the indicated 
rotation of molecule 1. b) Relaxed DFT structure obtained from the configuration in 
panel a).}
\end{figure}

Fig.~\ref{fig1}a) provides a schematic picture of the formation of Bjerrum defects in 
ice Ih. Defect-free ice Ih, in which each molecule is hydrogen-bonded to its four 
tetrahedrally positioned neighbors, obeys Pauling's two ice rules~\cite{Petrenko}: 
(i) each molecule offers/accepts two hydrogens to/from two neighboring molecules, and 
(ii) there is precisely one hydrogen between each nearest-neighbor pair of oxygens. 
The proton-disordered character implies that there is no long-range order in the 
orientations of the ${\rm H_2O}$ molecules or hydrogen bonds. The Bjerrum defect pair 
is a violation of ice rule (ii): it is obtained by the indicated rotation of molecule 
1, after which there are two hydrogens between molecules 1 and 2 (D defect) and none 
between molecules 1 and 3 (L defect). After the initial formation, the pair can be 
further separated by successive molecular rotations, eventually creating a pair of 
independent, or free, D and L defects. Their motion through the crystal provokes the 
rotation of the water molecules along their paths, providing a mechanism for 
electrical polarization. 

In both pure as well as doped ice, the conductivity $\sigma_{\rm DL}$ due to Bjerrum 
defects is essentially controlled by L defects~\cite{Petrenko}. According to 
Jaccard's electrical theory of ice~\cite{Jaccard1959} it takes the form
$\sigma_{\rm DL}=q_{\rm DL}\,n_{\rm L}\,\mu_{\rm L}$, 
where $q_{\rm DL}$ is the effective charge carried by the Bjerrum defects, $n_{\rm 
L}$ is the concentration of free L defects and $\mu_{\rm L}$ is their mobility. Its 
temperature dependence is described by~\cite{Petrenko}
\begin{equation}
\sigma_{\rm DL}(T)\sim\,\frac{q_{\rm DL}}{T}\exp(-E/k_B T) ,
\label{conduct}
\end{equation}
where $T$ is the absolute temperature, $k_B$ is Boltzmann's constant and $E$ is a 
characteristic activation energy. In {\em pure} ice, the latter is given 
by~\cite{Petrenko} 
\begin{equation}
E= {\textstyle \frac{1}{2}} E_{\rm DL}+E_{\rm Lm},
\label{Eact}
\end{equation}
where $E_{\rm DL}$ is the formation energy of an independent defect pair and $E_{\rm 
Lm}$ is the migration energy of a free L defect. 

While the activation energy can be determined experimentally by measuring  
$\sigma_{\rm DL} $ as a function of $T$ and adjusting it to Eq.~(\ref{conduct}), it 
is not possible to directly isolate the components $E_{\rm DL}$ and $E_{\rm Lm}$. To 
achieve this, additional conductivity measurements need to be carried out on {\em 
doped} ice samples,~\cite{Petrenko} in which an extrinsic concentration of L defects 
is injected. This indirect procedure, however, is subject to large uncertainties and 
has revealed incompatibilities between different experiments~\cite{Petrenko}.

Here we use a density-functional theory (DFT)~\cite{Martin} approach to explicitly 
compute the formation and migration energies $E_{\rm DL}$ and $E_{\rm Lm}$. Our 
calculations are performed using the 96-molecule supercell labeled $3\times2\times2$ 
in Ref.~\onlinecite{Hayward1997}. The calculations are executed using the VASP 
package~\cite{Kresse1993,Kresse1996a} using the Perdew-Wang 91 generalized-gradient 
approximation~\cite{Martin} and the projector-augmented-wave~\cite{Kresse1999} 
approach. Brillouin-zone sampling was limited to the $\Gamma$-point and we use a 
plane wave cut-off of $E_{\rm cut}=700$~eV. The effects of spurious image dipole 
interactions were evaluated~\cite{Kantorovich1999} and found to be negligible for all 
investigated structures.

First, we relax the defect-free crystal supercell, allowing both the atomic and 
supercell coordinates to relax at zero stress. The resulting hexagonal lattice 
parameters $a=4.383 \, \AA$ and $c=7.16\, \AA$, are $\sim 2$~\% below the 
experimental values measured at $T=10$~K~\cite{Petrenko}. The average intramolecular 
oxygen-hydrogen separation of $1.01\, \AA$ is in excellent agreement with the 
experimental value of $\sim1.006-1.008 \AA$. As appears typical of DFT calculations 
on ice Ih~\cite{Feibelman2002} our calculations slightly overestimate the sublimation 
energy of 0.69 eV compared to the experimental value 0.61 eV~\cite{Petrenko}.   

Next, we create an ``embryonic'' Bjerrum defect pair according to Fig.~\ref{fig1}a) 
and relax it at constant volume. The resulting structure is shown in 
Fig.~\ref{fig1}b) and is qualitatively similar to those observed in a recent 
molecular dynamics (MD) study~\cite{Grishina2004}. Compared to the initial geometry 
in Fig.~\ref{fig1}b), one can no longer recognize a D defect in the sense of its 
description in Fig.~\ref{fig1}a) due to the large electrostatic repulsion between the 
two hydrogen atoms~\cite{Petrenko,Podeszwa1999,Grishina2004}. The total-energy part 
of the formation energy of this structure is found to be 0.55 eV. To account for 
zero-point contributions, known to be relevant in ice~\cite{Petrenko}, we also 
evaluate the change in the local inter and intramolecular vibrational modes with 
respect to the defect-free crystal for the molecules in the vicinity of the defect, 
using the local harmonic approximation~\cite{LeSar1989}. Overall, the zero-point 
contribution lowers the formation energy by about 10\% to 0.50 eV. 

\begin{figure}
\includegraphics*[scale=0.7]{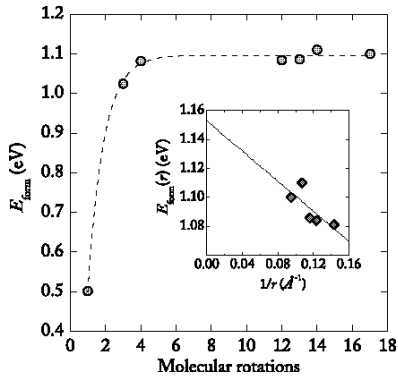}
\caption{\label{fig2} Formation energy of the Bjerrum defect pair as a function of 
the number of molecular rotations separating them. Dashed line is to guide the eye. 
Inset shows least-squares fit of the formation energies of the defect pairs separated 
by more than 3 molecular rotations to Eq.~(\ref{fit}).}
\end{figure}

To estimate the formation energy $E_{\rm DL}$ of an independent pair of Bjerrum 
defects, we move the L defect through the crystal by a series of molecular rotations, 
followed by structural relaxation. Assuming that the zero-point contribution is the 
same as for the embryonic defect pair, the formation energy as a function of the 
number of molecular rotations is shown in Fig.~\ref{fig2}. It quickly reaches a 
plateau value around 1.1 eV, subject to small fluctuations due to the disordered 
character of the hydrogen-bond network. However, a more accurate estimate of $E_{\rm 
DL}$ requires an analysis of its asymptotic behavior as a function of distance 
between the defect pair near the plateau value. Because of their effective charges 
${\textstyle \pm} q_{\rm DL}$, this behavior is expected to be of the form 
\begin{equation}
E_{\rm form}(r)=E_{\rm DL}-\frac{q^2_{\rm DL}}{4\, \pi \, \epsilon_0 
\,\epsilon_{\infty} r},
\label{fit}
\end{equation}
where $r$ is the distance between the point charges ${\textstyle \pm} q_{\rm DL}$, 
and $\epsilon_{\infty}$ is the high-frequency dielectric constant of ice 
Ih~\cite{Petrenko}. For each relaxed Bjerrum defect pair we position a point-charge 
$+q_{\rm DL}$ on the dangling proton of the D defect and center a charge $-q_{\rm 
DL}$ midway between the two oxygen atoms of the L defect, after which we adjust the 
formation energies of the defect pairs separated by more than 3 molecular rotations 
to Eq.~(\ref{fit}). The results, shown in the inset of Fig.~\ref{fig2}, are 
consistent with $1/r$ behavior and the intercept gives an asymptotic value $E_{\rm 
DL}=(1.153 \pm 0.04)$~eV. In addition, the slope of the fit provides an estimate for 
the effective charge which, using the experimental value 
$\epsilon_{\infty}=3.2$~\cite{Petrenko}, gives $q_{\rm DL} = (0.34 \pm 0.07)\,e$, in 
good agreement with the experimental value $q_{\rm DL}=\,0.38 \,e$~\cite{Petrenko}. 

\begin{figure}
\includegraphics*[scale=0.4]{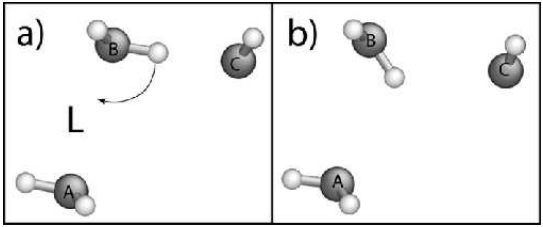}
\caption{\label{fig3} Identification of transition state for the migration of an 
independent L defect. a) Equilibrium structure of the independent L defect. b) 
Typical transition state.}
\end{figure}

Having determined $E_{\rm DL}$ we now compute the migration energy barrier $E_{\rm 
Lm}$. Given the disorder in the hydrogen-bond network, this barrier is expected to 
fluctuate depending on the local environment of the L defect. For this purpose, we 
computed 6 distinct barriers at 3 different L-defect sites (13, 14 and 17 molecular 
rotations, cf. Fig.~\ref{fig2}) by starting from the relaxed D-L configuration and 
rotating one of the 2 molecules hosting the L defect as shown in Fig.~\ref{fig3}a), 
constraining only the rotating proton while allowing full relaxation of all other 
coordinates. This yields transition states of the type shown in Fig.~\ref{fig3}b), in 
which the angle \angle ABC is essentially bisected by the coplanar rotating OH bond. 
The resulting energy barriers (neglecting zero-point effects) vary in the range 
$E_{\rm Lm}\simeq 0.10-0.14$~eV, indicating the significance of the disorder but an 
overall high mobility of free L defects. These results are consistent with the recent 
MD findings obtained at $T=230~K$~\cite{Podeszwa1999}, for which, assuming an attempt 
frequency $\nu_0=24$~THz typical for librational modes in ice Ih~\cite{Petrenko}, the 
present barriers give an average migration time ranging between 4 and 35 ps. 

\begin{figure}
\includegraphics*[scale=0.5]{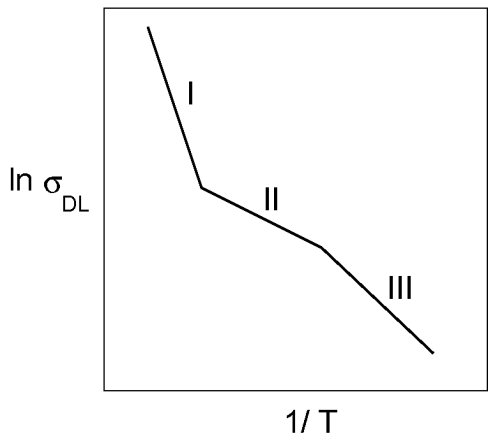}
\caption{\label{fig4} Schematic representation of a characteristic experimental 
Arrhenius plot of the conductivity $\sigma_{\rm DL}$ as a function of temperature in 
a doped ice Ih sample~\cite{Petrenko}.}
\end{figure}

The experimental estimates for the formation and migration energies are obtained from 
conductivity measurements on ice doped with a substitutional concentration of 
HF~\cite{Jaccard1959,Camplin1978} or HCl~\cite{Takei1987} molecules. Since each 
molecule has only one proton, they introduce an extrinsic, temperature-independent 
concentration of L defects. Measurements of the conductivity $\sigma_{\rm DL}$ as a 
function of $T$ then typically yield an Arrhenius-type plot with 3 distinct 
activation energies, as shown in Fig.~\ref{fig4}~\cite{Petrenko}. 

The high-temperature regime $I$ is believed to be controlled by intrinsic behavior, 
characterized by the activation energy Eq.~(\ref{Eact}). The lower-temperature 
regimes II and III are assumed to be dominated by the extrinsic L defects. In the 
former, the temperature dependence of the conductivity has been attributed to the 
{\em free} motion of extrinsic L defects~\cite{Petrenko} so that $E_{\rm II}=E_{\rm 
Lm}$. In regime III the temperature is so low that the extrinsic L defects are not 
completely dissociated from their dopant molecules so that the activation energy 
involves an additional dissociation energy, giving $E_{\rm III}={\textstyle 
\frac{1}{2}}E_{\rm diss}+E_{\rm Lm}$~\cite{Petrenko}.     

Considering the experimental results~\cite{Jaccard1959,Camplin1978,Takei1987} 
reproduced in Table~\ref{tab1}, the activation energy for regime I is quite well 
established, even for different dopant species, showing a dispersion of less than 
0.05 eV among the different experiments. The values for the individual components 
$E_{\rm DL}$ and $E_{\rm Lm}$, however, show considerably larger deviations. The 
migration barrier values vary between 0.19 eV and 0.315 eV, leading to formation 
energies $E_{\rm DL}$ ranging from 0.66 eV to 0.79 eV. 

Comparing these to our DFT calculations (cf. Table \ref{tab1}), we notice that both 
components of the activation energy deviate significantly from the experimental 
values. The DFT result for $E_{\rm DL}$ is more than 46~\% larger than the largest 
experimental value, whereas $E_{\rm Lm}$ is about 37~\% lower than the lowest 
experimental estimate. In this light, it is quite striking that, despite the 
discrepancies for the individual components, the DFT estimate for the net activation 
energy in Eq.~(\ref{Eact}) agrees quite well with the experimental results, deviating 
about 10~\% from the highest experimental estimate. This seems to demonstrate the 
difficulty involved in the interpretation of conductivity experiments in doped ice 
samples under conditions not controlled by intrinsic properties. Specifically, the 
fact that the DFT estimate for the migration barrier is systematically and 
significantly lower than all experimental estimates indicates that, as suggested in 
Ref.~\onlinecite{Petrenko}, the regime interpreted as being controlled by free 
extrinsic L defects, may in fact involve the activity of traps that obstruct their 
motion, leading to the higher effective migration barriers deduced experimentally. 

In summary, we have conducted a first-principles study of the energetics of Bjerrum 
defects in ice Ih and compared the results to experimental electrical conductivity 
for doped ice samples. The results provide new insight into the parameters in 
Jaccard´s microscopic electrical theory of ice. While the DFT value for the net 
activation energy is in good agreement with experiment, we find that its two 
components have quite different values from those inferred from experiment. In 
particular, our results predict a migration barrier for L-defect motion that is 
significantly lower than the lowest experimental estimate, hinting at the presence of 
traps in the regime usually interpreted as being controlled by the free migration of 
extrinsic L defects.
 
\begin{acknowledgments}
The authors gratefully acknowledge financial support from the Brazilian agencies 
FAPESP and CNPq.
\end{acknowledgments}


\newpage

\begin{table}
\caption{\label{tab1} Activation energies and their components $E_{\rm DL}$ and 
$E_{\rm Lm}$ determined from doped ice experiments and the present DFT calculations.}
\centering
\begin{tabular}{l|l|l|l}
\cline{1-4}
\vbox to4.04ex{\vspace{2pt}\vfil\hbox to18.80ex{\hfil \hfil}} & 
\vbox to4.04ex{\vspace{2pt}\vfil\hbox to12.80ex{\hfil $E$\hfil}} & 
\vbox to4.04ex{\vspace{2pt}\vfil\hbox to12.80ex{\hfil $E_{\rm Lm}$ \hfil}} & 
\vbox to4.04ex{\vspace{2pt}\vfil\hbox to12.80ex{\hfil $E_{\rm DL}$ \hfil}} \\
\cline{1-4}
\cline{1-4}
\vbox to2.04ex{\vspace{2pt}\vfil\hbox to18.80ex{\hfil Ref.~\onlinecite{Jaccard1959} 
(HF)\hfil}} & 
\vbox to2.04ex{\vspace{2pt}\vfil\hbox to12.80ex{\hfil 0.575\hfil}} & 
\vbox to2.04ex{\vspace{2pt}\vfil\hbox to12.80ex{\hfil 0.235\hfil}} &
\vbox to2.04ex{\vspace{2pt}\vfil\hbox to12.80ex{\hfil 0.68\hfil}} \\

\cline{1-4}
\vbox to2.04ex{\vspace{2pt}\vfil\hbox to18.80ex{\hfil Ref.~\onlinecite{Camplin1978} 
(HF)\hfil}} & 
\vbox to2.04ex{\vspace{2pt}\vfil\hbox to12.80ex{\hfil 0.625\hfil}} & 
\vbox to2.04ex{\vspace{2pt}\vfil\hbox to12.80ex{\hfil 0.315\hfil}} &
\vbox to2.04ex{\vspace{2pt}\vfil\hbox to12.80ex{\hfil 0.664\hfil}}\\

\cline{1-4}
\vbox to2.04ex{\vspace{2pt}\vfil\hbox to18.80ex{\hfil Ref.~\onlinecite{Takei1987} 
(HCl)\hfil}} & 
\vbox to2.04ex{\vspace{2pt}\vfil\hbox to12.80ex{\hfil 0.585 \hfil}} & 
\vbox to2.04ex{\vspace{2pt}\vfil\hbox to12.80ex{\hfil 0.190 \hfil}} &
\vbox to2.04ex{\vspace{2pt}\vfil\hbox to12.80ex{\hfil 0.79  \hfil}} \\

\cline{1-4}
\vbox to2.04ex{\vspace{2pt}\vfil\hbox to18.80ex{\hfil This work \hfil}} & 
\vbox to2.04ex{\vspace{2pt}\vfil\hbox to12.80ex{\hfil $0.696 \pm 0.04$ \hfil}} & 
\vbox to2.04ex{\vspace{2pt}\vfil\hbox to12.80ex{\hfil $0.120 \pm 0.02$ \hfil}} &
\vbox to2.04ex{\vspace{2pt}\vfil\hbox to12.80ex{\hfil $1.153 \pm 0.04$ \hfil}} \\

\cline{1-4}
\end{tabular}
\end{table}

\pagebreak

\newpage

\newpage

\newpage

\end{document}